\documentstyle[aps,12pt,preprint,epsfig]{revtex}
 
\draft
\begin{document}
 
\title{$\chi_{cJ}$ polarizations at the Fermilab Tevatron}
\author{Feng Yuan}
\address{\small {\it Department of Physics, Peking University, Beijing
100871, People's Republic of China\\
Institut f\"ur Theoretische Physik der Universit\"at, Philosophenweg 19, 
D-69120 Heidelberg, Germany}}
\author{Kuang-Ta Chao}
\address{\small {\it China Center of Advanced Science and Technology
(World Laboratory), Beijing 100080, People's Republic of China\\
and Department of Physics, Peking University, Beijing 100871,
People's Republic of China}}
\maketitle
 
\begin{abstract}
We propose the measurement of $\chi_{cJ}$ polarizations at high energy
hadron colliders to study heavy quarkonium production
mechanism. We find that the color-singlet model in the $k_t$
factorization approach predicts very different behavior for $\chi_{cJ}$
polarizations at the Fermilab Tevatron compared with the NRQCD predictions
in the collinear parton model. In the color-singlet $k_t$ factorization
approach, for both $\chi_{c1}$ and $\chi_{c2}$ productions, the helicity 
$h=0$ states dominate over other helicity states at large $p_T$.
These properties are very useful in 
distinguishing between the two production mechanisms which are related
to the interesting issue of $J\psi$ and $\psi'$ polarizations,
and may provide a crucial test for the $k_t$ factorization
approach.
\end{abstract}
 
Studies of heavy quarkonium production in high energy collisions provide
important information on both perturbative and nonperturbative QCD.
In recent years, heavy quarkonium production has attracted much attention
from both theory and experiment.
To explain the $J/\psi$ and $\psi'$ surplus problem
of large transverse momentum production at Tevatron\cite{fa},
the color-octet production mechanism was introduced for the
description of heavy quarkonium production\cite{surplus} based
on the NRQCD factorization framework\cite{nrqcd}.
During the last few years, extensive studies have been performed
for the test of this color-octet production mechanism.
However, most recently the CDF collaboration have reported their
preliminary measurements on the polarizations of the promptly
produced charmonium states\cite{pola}, which appear not to
support the color-octet predictions that the directly produced $S$-wave
quarkonia have transverse polarizations at large $p_T$
\cite{th-pola,beneke}
In \cite{braaten,kniel}, the authors considered the feeddown contributions from
$\chi_c$ decays, and found the prompt $J/\psi$ polarization disagree with 
the CDF data at large $p_T$ by 3 standard deviations.
This conflict shows that the heavy quarkonium production mechanism may be
more complicated than we knew before, and further studies
on heavy quarkonium production mechanisms other than the color-octet
mechanism in the collinear parton model are still needed at present.
 
In\cite{teryaev}, the authors studied 
$\chi_{cJ}$ hadroproduction at Tevatron in the $k_t$ factorization
approach\cite{ca,co}. Their results show that only color-singlet
contributions can reproduce the Tevatron data on $\chi_{cJ}$
production in the $k_t$ factorization approach,
and the color-octet contributions disagree with the data.
However, we note that previous studies in the collinear parton model
have also given a good description for the data of $\chi_{cJ}$ production
at Tevatron in the NRQCD approach including both color-singlet and 
color-octet contributions\cite{cho}.
In this context, we have two different mechanisms, i.e., the color-singlet
$k_t$ factorization approach and the NRQCD approach in the collinear
parton
model, of which both can successfully describe the Tevatron data on the
large $p_T$ $\chi_{cJ}$ production rates.
So, it is quite urgent now to distinguish between these two mechanisms 
for understanding 
heavy quarkonium production at high energy hadron colliders.
For this purpose, we propose here the measurements of $\chi_{cJ}$ 
polarizations at the Fermilab Tevatron.
From  our calculations we find that the color-singlet model in the $k_t$
factorization approach predicts very different behavior for $\chi_{cJ}$
polarizations compared with the NRQCD predictions
in the collinear parton model.
In the color-singlet $k_t$ factorization approach, 
for both $\chi_{c1}$ and $\chi_{c2}$ productions, {\it the helicity $h=0$ 
states dominate over other helicity states at large transverse momentum}.
This novel property can provide a crucial test for this production
mechanism.
Furthermore, the $\chi_{cJ}$ polarization measurements can help to clarify
the present conflict between the color-octet predictions and the
experimental 
data on prompt $J/\psi$ and $\psi'$ polarizations at Tevatron.
 
The polarized cross section formulas for $\chi_{cJ}$ hadroproduction
in the NRQCD approach have been calculated in \cite{wise,cho,kniel} 
(including both color-singlet and color-octet processes).
In this paper, we will calculate the color-singlet $\chi_{cJ}$ polarized
cross sections in the $k_t$ factorization approach. We will not include
the color-octet processes in this approach, because their contributions to
$\chi_{cJ}$ production disagree with the Tevatron data in
shape\cite{teryaev}.
 
The $k_t$-factorization approach differs greatly from the
conventional collinear approximation because it takes the non-vanishing
transverse momenta of the scattering partons into account.
The conventional gluon densities are replaced by the unintegrated gluon
distributions which depend on the transverse momentum $k_t$.
In the calculations, for every 4-momenta $k_i$
we make a Sudakov decomposition as
\begin{equation}
k_i=\alpha_i p_1+\beta_i p_2+\vec{k}_{iT},
\end{equation}
where $p_1$ and $p_2$ are the momenta of the incoming hadrons.
In the high energy limit, we have $p_1^2=0$, $p_2^2=0$, and
$2p_1\cdot p_2=s$, where $s$ is the c.m. energy squared.
$\alpha_i$ and $\beta_i$ are the momentum fractions of $p_1$ and $p_2$
respectively.
$k_{iT}$ is the transverse momentum, which satisfies
\begin{equation}
k_{iT}\cdot p_1=0,~~~
k_{iT}\cdot p_2=0.
\end{equation}
For the momenta of the incident gluons $q_1$ and $q_2$, we
have the following decompositions\cite{co},
\begin{equation}
q_1=x_1 p_1+q_{1T},~~~~q_2=x_2 p_2+q_{2T}.
\end{equation}
That is to say, the longitudinal component of $q_1$ ($q_2$)
is only in the direction of light-like vector $p_1$ ($p_2$).
 
Using the above defined Sudakov variables, we can express the
polarized cross sections for $\chi_{cJ}$ hadroproduction
as the following form,
\begin{equation}
\label{xs}
d\sigma(p\bar p\rightarrow \chi_{cJ}^{(\lambda)} X)=
\frac{1}{64\times 16\pi}\frac{d\alpha_\chi }{\alpha_\chi}
d^2q_{1T}d^2q_{2T}\frac{f(x_1;q_{1T}^2)}{q_{1T}^2}
\frac{f(x_2;q_{2T}^2)}{q_{2T}^2}\frac{|A_0^{(\lambda)}(q_{1T},q_{2T})|^2}
{q_{1T}^2q_{2T}^2},
\end{equation}
where $\lambda$ denotes the helicity of $\chi_{cJ}$,
and $\alpha_\chi$ is the momentum fraction of $p_1$ carried by
$\chi_{cJ}$.
The $\chi_{cJ}$
tranverse momentum $p_T$ comes from the sum of the transverse momenta of
$q_1$
and $q_2$ as, $\vec{p}_T=\vec{q}_{1T}+\vec{q}_{2T}$.
The amplitude squared $|A_0^{(\lambda)}|^2$ describes $\chi_{cJ}$
(with helicity $\lambda$) production in the gluon-gluon
fusion processes $g+g\rightarrow \chi_{cJ}^{(\lambda)}$. 
To calculate these helicity amplitudes, we need the polarization sums 
for individual helicity
levels of $\chi_{cJ}$. For $\chi_{c1}$, the longitudinal and transverse
polarization sums can be written in the following covariant forms
\begin{eqnarray}
\sum\limits_{\lambda=0}\epsilon_\alpha^{(\lambda)}\epsilon_\beta^{(\lambda)*}
&=&P_{\alpha\beta}^L,\\
\sum\limits_{|\lambda|=1}\epsilon_\alpha^{(\lambda)}\epsilon_\beta^{(\lambda)*}
&=&P_{\alpha\beta}^T=P_{\alpha\beta}-P_{\alpha\beta}^L,
\end{eqnarray}
where 
\begin{equation}
P_{\alpha\beta}=-g_{\alpha\beta}+\frac{p_\alpha p_\beta}{p^2}.
\end{equation}
And in the laboratory frame (the helicity frame), 
$P_{\alpha\beta}^L$ is expressed
as\cite{beneke}
\begin{equation}
P_{\alpha\beta}^L=\frac{(p\cdot Q)^2}{(p\cdot Q)^2-M^2s}
( \frac{p_\alpha}{M}-\frac{M}{p\cdot Q}Q_\alpha )
( \frac{p_\beta}{M}-\frac{M}{p\cdot Q}Q_\beta ),
\end{equation}
where $M=2m_c$ is the mass of $\chi_{cJ}$, and $Q=p_1+p_2$ is the sum of
the initial hadron 4-momenta.
For $\chi_{c2}$, the polarization sums for individual helicity levels 
($\lambda=0,~1,~2$) can also be expressed in terms of $P_{\alpha\beta}$,
$P_{\alpha\beta}^T$, and $P_{\alpha\beta}^L$\cite{wise}.
With these polarization sums, we can calculate the production cross
sections
for individual helicity states of $\chi_{cJ}$, which are more 
involved and will be presented elsewhere.
We have checked that these cross section formulas can numerically
reproduce
the results of \cite{teryaev} for the inclusive production rates of 
$\chi_{cJ}$ at Tevatron after summing up all helicity states
contributions.
 
The $\chi_{cJ}$ polarizations can be measured by studying the photon's
angular
distribution in the $\chi_{cJ}$ rest frame in the decay processes 
$\chi_{cJ}\rightarrow J/\psi\gamma$.
These angular distributions have the following form,
\begin{equation}
\frac{d\Gamma(\chi_{cJ}\rightarrow J/\psi\gamma)}{d\cos \theta}\propto
\frac{3}{2(3+\alpha)}(1+\alpha \cos^2\theta),
\end{equation}
where $\theta$ is the angle between the photon's 3-momentum in $\chi_{cJ}$
rest frame and the $\chi_{cJ}$ 3-momentum in the laboratory frame.
$\alpha$ is the polarization parameter (angular distribution parameter).
For $\chi_{c1}$, $\alpha$ is defined as\cite{wise}
\begin{equation}
\alpha=\frac{2-3\rho}{2+\rho},
\end{equation}
where 
\begin{equation}
\rho=\frac{d\sigma(\chi_{c1}^{(|\lambda|=1)})}
{d\sigma(\chi_{c1})}.
\end{equation}
For $\chi_{c2}$, the polarization parameter is\cite{wise}
\begin{equation}
\alpha=-\frac{6-3\eta-12\tau}{10-\eta-4\tau},
\end{equation}
where 
\begin{equation}
\eta=\frac{d\sigma(\chi_{c2}^{(|\lambda|=1)})}
{d\sigma(\chi_{c2})},~~
\tau=\frac{d\sigma(\chi_{c2}^{(|\lambda|=2)})}
{d\sigma(\chi_{c2})}.
\end{equation}
 
For numerical calculations, we choose the unintegrated gluon distribution
of \cite{martin} which can well fit the $F_2(x,Q^2)$ data over a wide
range
of $x$ and $Q^2$, and 
 we set the scales $\mu^2$ for the
strong coupling constant $\alpha_s(\mu^2)$ in the
amplitude squared $|A_0|^2$ to be $q_{1T}^2$ for the interaction
vertex associated with the incident gluon $q_1$, and 
$q_{2T}^2$ for the vertex associated with $q_2$\cite{levin,teryaev}.
 
We first display in Fig.~1 the production ratio of $\chi_{c1}$ to
$\chi_{c2}$
at the Tevatron as a function of $p_T$, 
$R=\sigma(\chi_{c1})/\sigma(\chi_{c2})$.
The solid line is for the color-singlet prediction in the $k_t$
factorization
approach, and the dotted-dashed line for the NRQCD prediction in the
collinear
parton model. For comparison, in this figure we also plot the results for 
other two cases in the collinear parton model:
the color-singlet prediction as the dotted line and the color-octet 
prediction as the dashed line.
However, we must note that neither the color-singlet contributions
nor the color-octet contributions alone can describe the Tevatron data
on $\chi_{cJ}$ productions\cite{cho}, and in this collinear parton model 
only the NRQCD predictions (including both
the color-singlet and the color-octet contributions) can make sense
to describe the Tevatron data on $\chi_{cJ}$ productions.
From Fig.~1, we can see that the $R$ ratio increases as $p_T$ increases
in 
the color-singlet model $k_t$ factorization approach, and its value
approaches 
to $2.0$ at large transverse momentum, which means that at large $p_T$
$\chi_{c1}$ production dominates over $\chi_{c2}$ production.
In contrast, the NRQCD approach predicts the $R$ ratio to be much smaller, 
and its
value approaches to $0.6$ at large $p_T$. This is because, at large 
transverse momentum the color-octet gluon fragmentation (the
${}^3S_1^{(8)}$ 
channel) dominates the
$\chi_{cJ}$ productions in NRQCD, which leads to $\chi_{cJ}$ production
rates as $\sigma(\chi_{c0}):\sigma(\chi_{c1}):\sigma(\chi_{c2})=1:3:5$
(consistent with our numerical calculations at large $p_T$).
This figure shows that the difference on $R$ ratio between the
color-singlet 
$k_t$ factorization approach and the NRQCD approach in the collinear
parton model
is distinctive at sufficiently large $p_T$.
 
We then study the $\chi_{cJ}$ polarizations at Tevatron. With the
polarized 
cross section formulas, we can calculate the production rates for definite
helicity states of $\chi_{cJ}$, and get the angular distribution parameter 
$\alpha$. From our numerical calculations, we find that the color-singlet
$k_t$ factorization approach predicts very different behavior for
$\chi_{cJ}$
polarizations at the Fermilab Tevatron compared with the NRQCD predictions
in the collinear parton model.
In the color-singlet $k_t$ factorization approach, 
for both $\chi_{c1}$ and $\chi_{c2}$ productions, {\it the helicity $h=0$ 
states dominate over other helicity states at large transverse momentum}.
This property has distinguished consequence to the decay angular
distribution
parameter $\alpha$ for $\chi_{cJ}\rightarrow J/\psi\gamma$ processes.
These results are displayed in Figs.~2-4.
For $\chi_{c1}\rightarrow J/\psi\gamma$, at large $p_T$ the color-singlet
$k_t$ factorization approach predicts $\alpha$ around $0.8$ while the
NRQCD
approach in the collinear parton model predicts $\alpha$ around $0.2$.
For  $\chi_{c2}\rightarrow J/\psi\gamma$, the difference between these two
mechanisms are more distinctive. The color-singlet $k_t$ factorization 
approach predicts  $\alpha(\chi_{c2}\rightarrow J/\psi\gamma)$ to be
negative (down to $-0.6$) at large $p_T$, 
while the NRQCD approach in the collinear
parton model predicts $\alpha$ to be positive (around $0.3$).
 
For the experimental measurement, it may be difficult to distinguish
between
$\chi_{c1}$ and $\chi_{c2}$ contributions in the observation of the 
photon's angular 
distributions in the decay processes $\chi_{cJ}\rightarrow J/\psi\gamma$.
So, it may be more useful to give the angular distributions in 
 $\chi_{cJ}\rightarrow J/\psi\gamma$ with both $\chi_{c1}$ and $\chi_{c2}$
 taken into account. We plot this result in Fig.~4.
From this figure, we find that the color-singlet $k_t$ factorization 
approach predicts $\alpha$ for $\chi_{cJ}\rightarrow J/\psi\gamma$ 
being about $0.5$ while the NRQCD approach in the collinear parton model
predicts $\alpha$ around $0.25$ at large transverse momentum.
The difference between these two mechanisms is still distinctive.
 
Finally, we note that the polarized cross section formulas for $\chi_{cJ}$
production can also be used to predict the polarization of $J/\psi$
which comes from $\chi_{cJ}$ feeddown decays. $J/\psi$ polarization can be
measured by the lepton's angular distribution in the $J/\psi$ rest frame
in $J/\psi\rightarrow \mu^+\mu^-$ decay process. These distributions have 
a similar form to those for $\chi_{cJ}$ decays,
\begin{equation}
\frac{d\Gamma(J/\psi\rightarrow \mu^+\mu^-)}{d\cos \theta}\propto
\frac{3}{2(3+\alpha)}(1+\alpha \cos^2\theta),
\end{equation}
where $\theta$ is the angle between the 3-momentum of the lepton in
$J/\psi$
rest frame and the 3-momentum of $J/\psi$ in the laboratory frame.
$\alpha$ is the polarization parameter, and is equal to 
\begin{equation}
\alpha=\frac{3\xi-2}{2-\xi},
\end{equation}
where $\xi$ is the ratio of the transversely polarized to the total
$J/\psi$,
which
can be calculated by using the polarized cross sections for $\chi_{cJ}$
production\cite{wise}.
In Fig.~5 we give the $J/\psi$ polarization from the feeddown 
contributions of $\chi_{cJ}$ decays. Again, we find that 
the color-singlet $k_t$ factorization 
approach predicts $\alpha (J/\psi\rightarrow \mu^+\mu^-)$ 
being about $0.5$ while the NRQCD approach in the collinear parton model
predicts $\alpha$ around $0.25$ at large $p_T$.

At the Tevatron, the CDF collaboration has measured the inclusive
production cross sections of $\chi_{cJ}$ states, which contribute about
$30\%$ of prompt $J/\psi$ production in a wide range of $p_T$\cite{fa}.
The $\chi_{cJ}$ states can be identified with a photon plus the $J/\psi$ 
which decays into a muon pair. The production cross section of $\chi_{cJ}$
is found to be comparable to that of $J/\psi$, and is not small.
Unfortunately, the present statistics of Tevatron Run I is not high enough 
to allow separate 
polarization measurements for this part of $J/\psi$ (from $\chi_c$ decays) 
from the direct $J/\psi$ production and $\psi'$ decay's contribution
\cite{pola}.
However, with the upgrade Tevatron Run II, the luminosity will be
increased
by a factor of more than 8, we will have much more data for  
$J/\psi$ and $\chi_{cJ}$ production,  
so it will be feasible to
distinguish between different contributions to prompt $J/\psi$
polarizations,
and then to measure the $\chi_{cJ}$ polarizations.

In conclusion, in this paper we have calculated the $\chi_{cJ}$ 
polarizations at high energy hadron colliders in the color-singlet
$k_t$ factorization approach and the NRQCD approach in the collinear 
parton model. We find the difference on the polarization parameters
for $\chi_{cJ}$ and their decay products 
$J/\psi$ between these two approaches 
are distinctive at large transverse momentum.
Therefore, $\chi_{cJ}$ polarizations can be used to study these
two production mechanisms at hadron colliders and may provide
important information on heavy quarkonium polarization mechanisms.
Especially, the color-singlet $k_t$ factorization approach predicts that
$\chi_{cJ}$ productions are dominated by the helicity $h=0$ states
at large transverse momentum. This unique property may provide 
a crucial test for this production mechanism.
 
\acknowledgments
This work was supported in part by the National Natural Science Foundation
of China, the Ministry of Education of China, and the State
Commission of Science and Technology of China.

\vskip 10mm
\centerline{\bf \large Figure Captions}
\vskip 1cm
\noindent
FIG. 1. The production ratio of $\chi_{c1}$ to $\chi_{c2}$ as a function 
of $p_T$. The solid line is for the color-singlet $k_t$ factorization 
approach prediction, the dotted-dashed line for the NRQCD prediction in
the
collinear parton model, the dotted and dashed lines are respectively for
the
 color-singlet and color-octet predictions alone 
in the collinear parton model.
 
\noindent
FIG. 2. The polarization parameter $\alpha$ for $\chi_{c1}\rightarrow
J/\psi
\gamma$ as a function of $p_T$. The definitions of the curves are the same
as in Fig.~1.

\noindent
FIG. 3. The same as in Fig.~2 but for $\chi_{c2}$.

\noindent
FIG. 4. The same as in Fig.~2 but for $\chi_{cJ}\rightarrow J/\psi
\gamma$ with both $\chi_{c1}$ and $\chi_{c2}$ taken into account.
 
\noindent
FIG. 5. The polarization parameter $\alpha$ for $J/\psi\rightarrow \mu^+
\mu^-$ for $J/\psi$ coming from $\chi_{cJ}$ decays.
 
\begin{figure}[thb]
\begin{center}
\epsfig{file=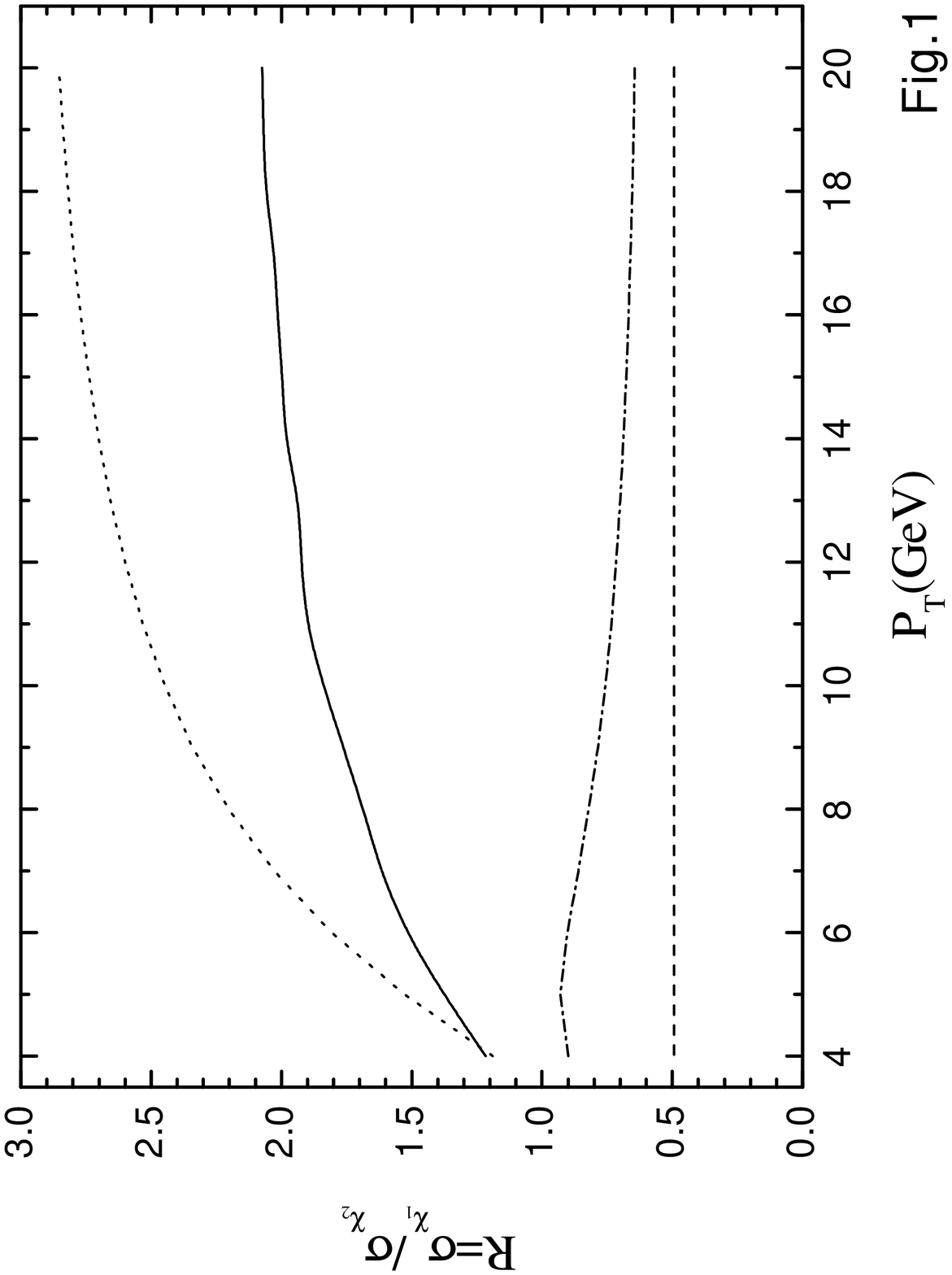,angle=0,width=16cm,height=18cm}
\end{center}
\end{figure}

\begin{figure}[thb]
\begin{center}
\epsfig{file=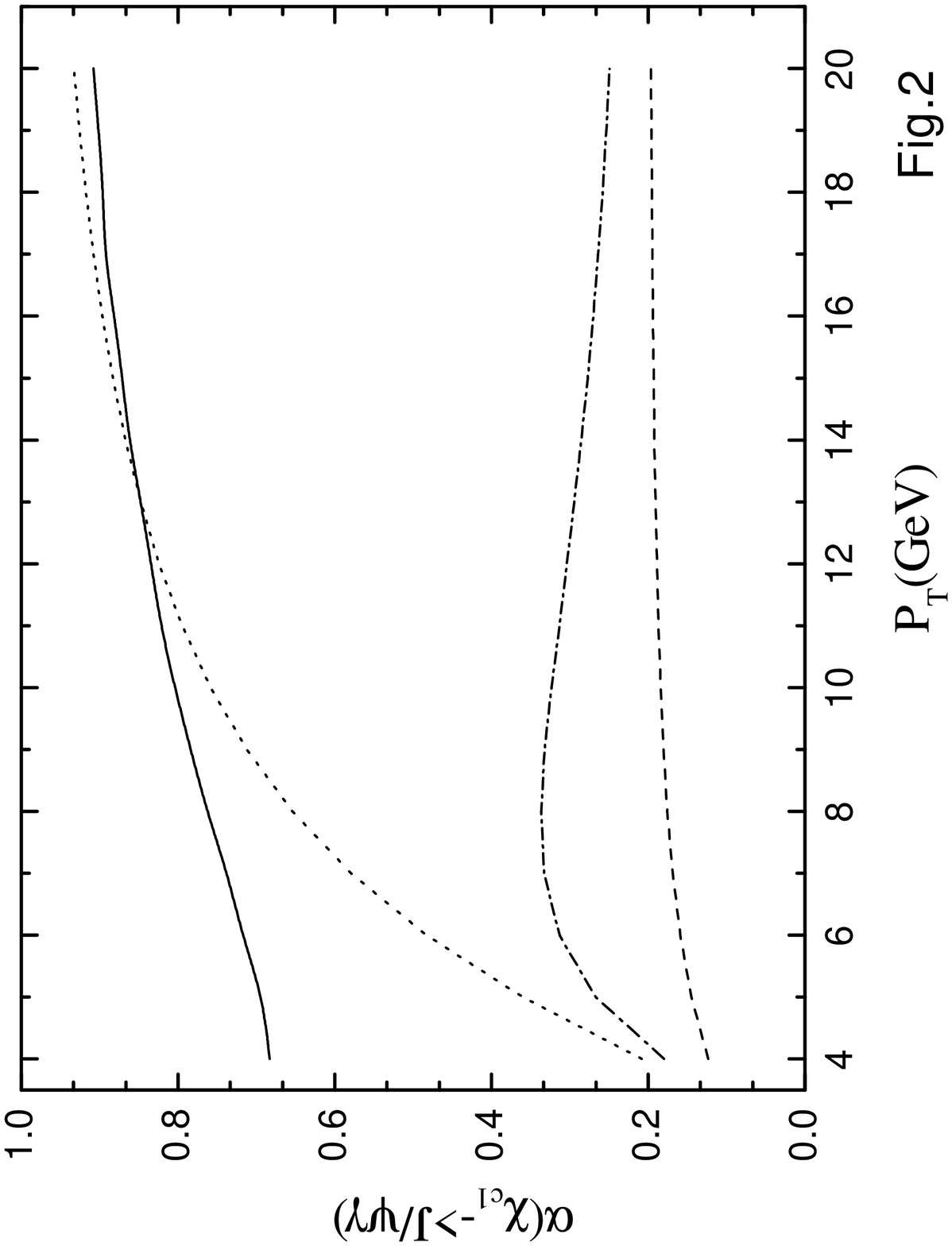,angle=0,width=16cm}
\end{center}
\end{figure}

\begin{figure}[thb]
\begin{center}
\epsfig{file=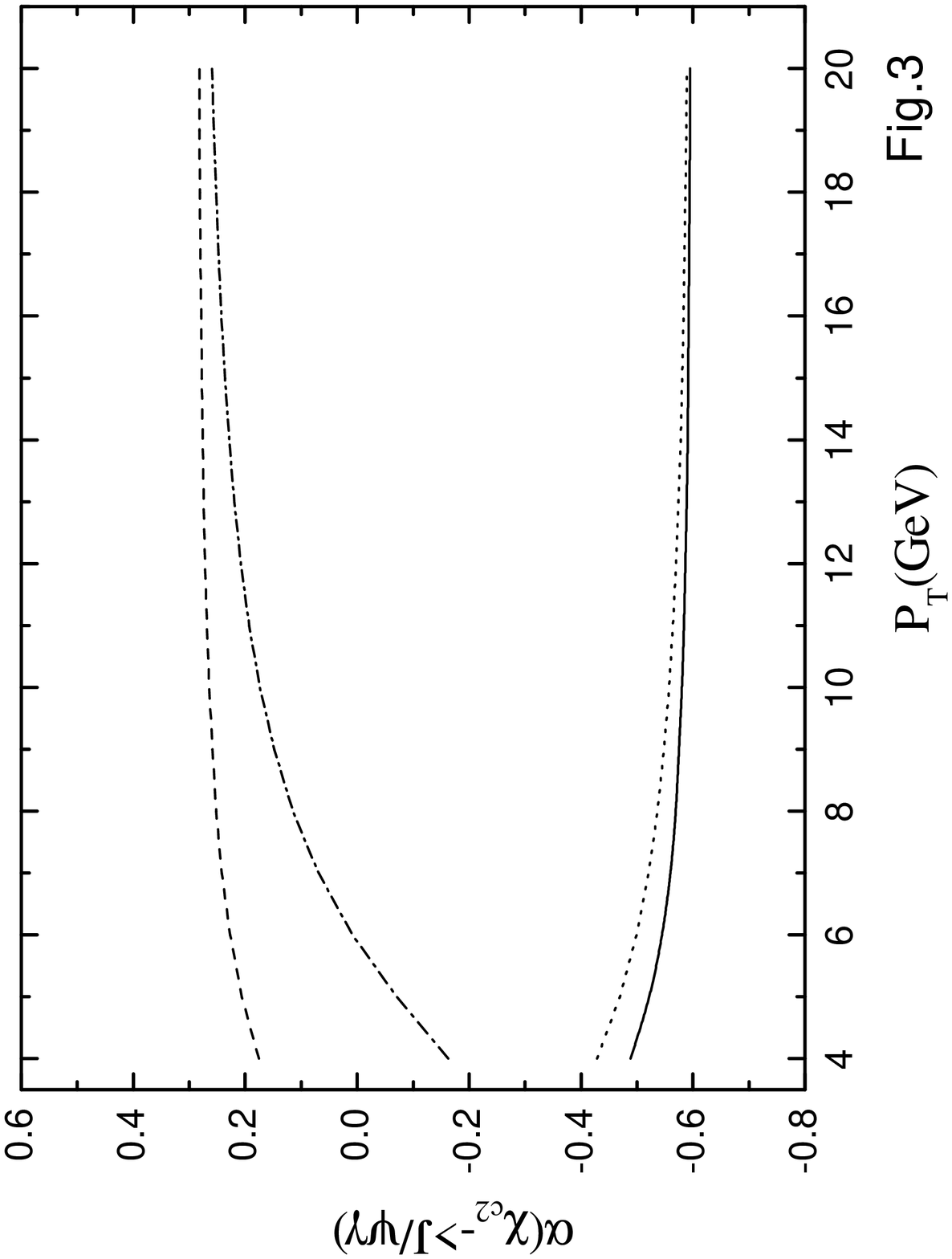,angle=0,width=16cm}
\end{center}
\end{figure}

 \begin{figure}[thb]
\begin{center}
\epsfig{file=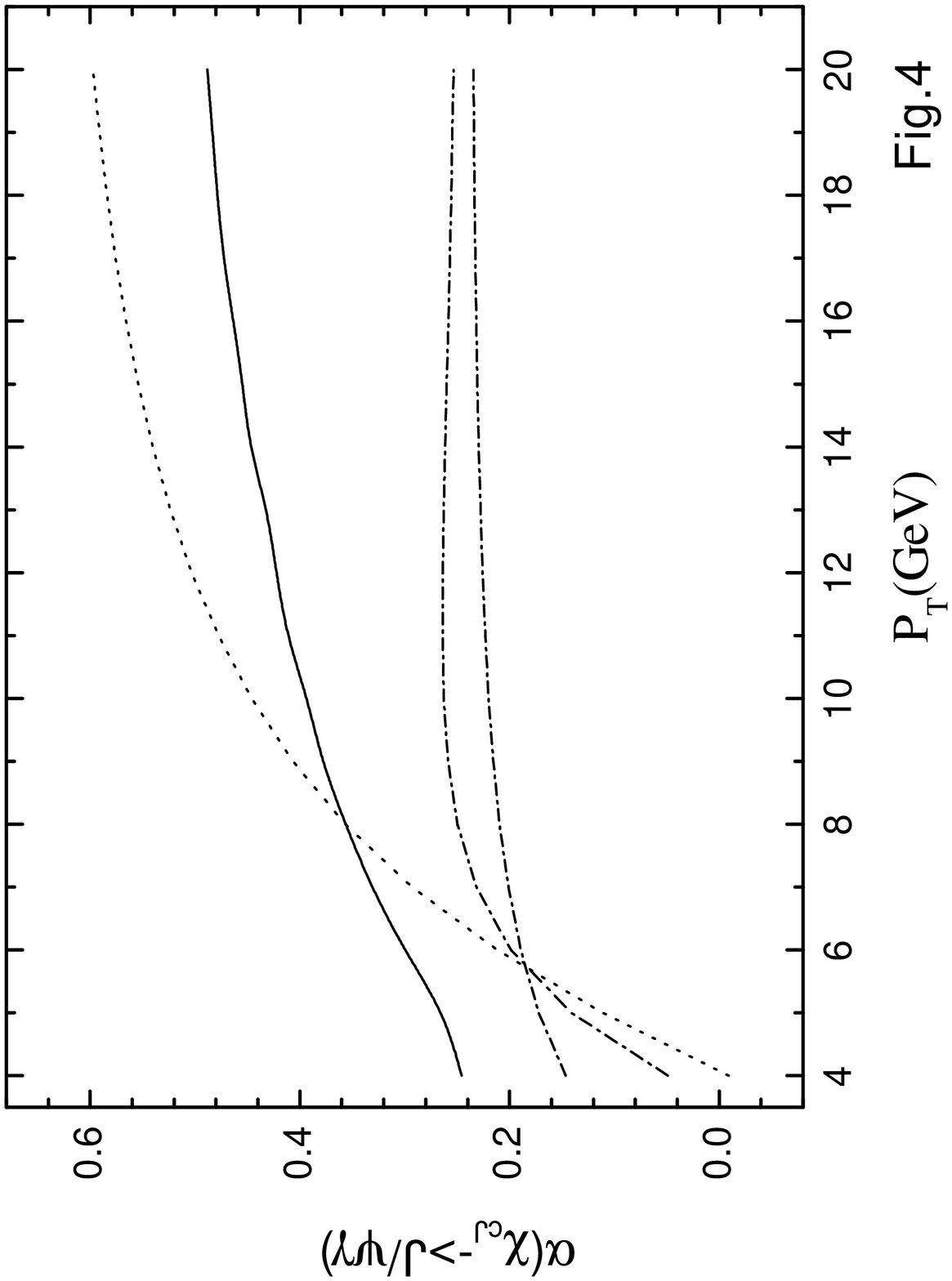,angle=0,width=16cm}
\end{center}
\end{figure}

\begin{figure}[thb]
\begin{center}
\epsfig{file=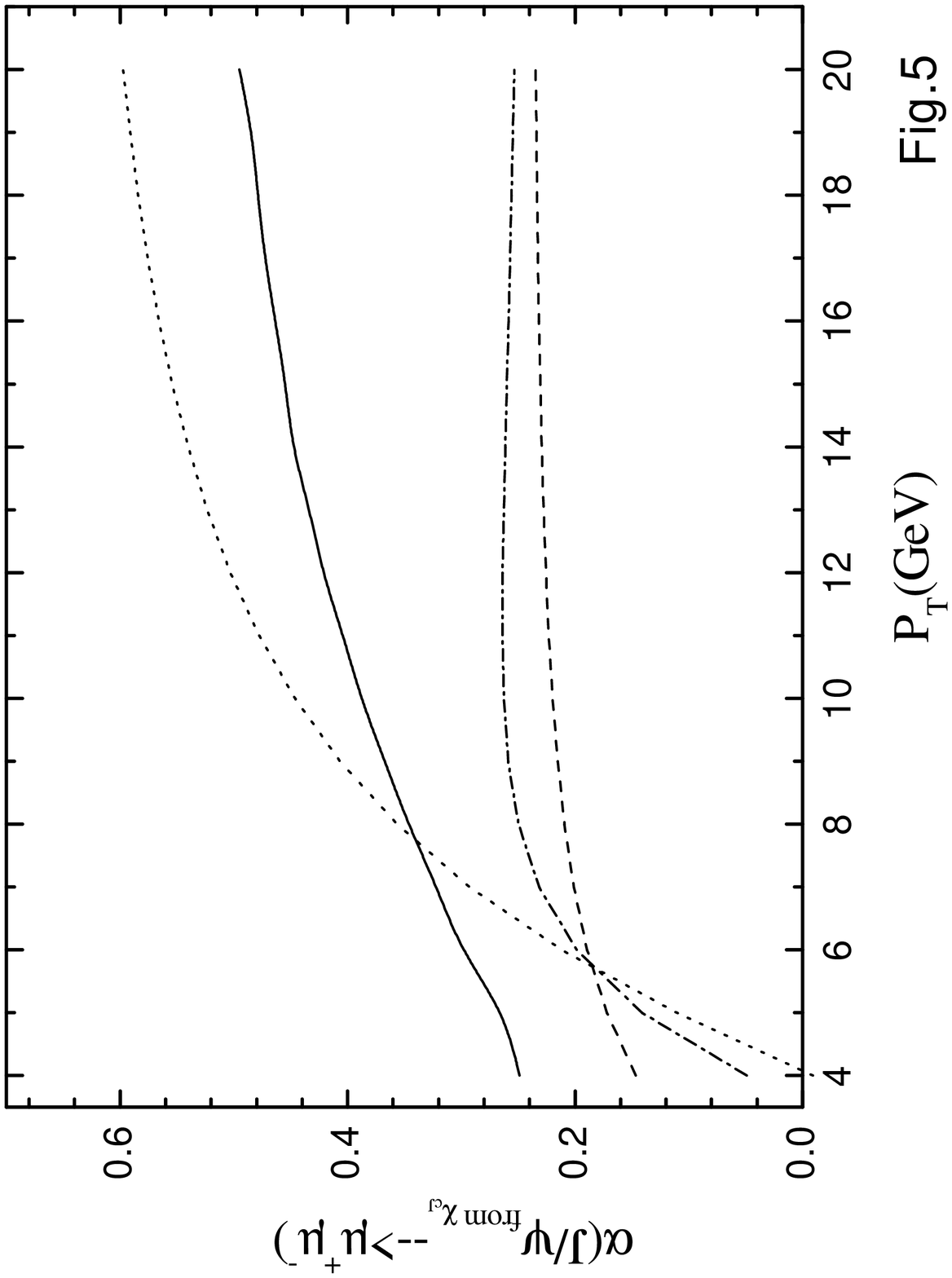,angle=0,width=16cm}
\end{center}
\end{figure}

\end{document}